\documentclass{aa}  

\usepackage{graphicx}
\usepackage{enumitem}
\usepackage{multirow}
\usepackage{booktabs}
\usepackage{array}
\usepackage{amsmath}
\usepackage{ragged2e}
\newcolumntype{L}[1]{>{\RaggedRight\arraybackslash}p{#1}}
\usepackage{txfonts}
\usepackage{xcolor}
\usepackage{comment}
\usepackage{subfig}
\usepackage[colorlinks, citecolor=blue, linkcolor=blue]{hyperref}
\usepackage[normalem]{ulem}

\providecommand{\gaia}{\textit{Gaia }}

\providecommand{\lmc}{$\mathnormal{G}_{\text{LMC}}$ }
\providecommand{\smc}{$\mathnormal{G}_{\text{SMC}}$ }
\providecommand{\mw}{$\mathnormal{G}_{\text{MW}}$ }
\providecommand{\lmcnospace}{$\mathnormal{G}_{\text{LMC}}$}
\providecommand{\smcnospace}{$\mathnormal{G}_{\text{SMC}}$}
\providecommand{\mwnospace}{$\mathnormal{G}_{\text{MW}}$}

\newcommand{\vphires}{V_\phi- \overline{V_\phi}}

\begin{document} 

   \title{Asymmetries in the LMC velocity maps}

   \subtitle{}

   \author{M. Sch\"olch \inst{1,2,3}$^,$\thanks{Email: marie.schoelch@fqa.ub.edu}
        \and Ó. Jiménez-Arranz \inst{4,2}
        \and M. Romero-Gómez\inst{1,2,3}
        \and X. Luri\inst{1,2,3}
        \and \\ D. Hobbs \inst{4}
        \and D. Salmerón-Larraz \inst{1}
        \and M. López Vilamajó \inst{1}
        }

    \authorrunning{M. Sch\"olch et al.}

    \institute{Departament de Física Quàntica i Astrofísica (FQA), Universitat de Barcelona (UB), C Martí i Franquès, 1, 08028 Barcelona, Spain 
    \and
    Institut de Ciències del Cosmos (ICCUB), Universitat de Barcelona, Martí i Franquès 1, 08028 Barcelona, Spain 
    \and
    Institut d’Estudis Espacials de Catalunya (IEEC), C Gran Capità, 2-4, 08034 Barcelona, Spain
    \and
    Lund Observatory, Division of Astrophysics, Department of Physics, Lund University, Box 43, SE-22100, Lund, Sweden}

   \date{Received Month XX, 2025; accepted Month XX, 2025}

  \abstract
  {The analysis of precise \gaia DR3 astrometry in the LMC region has revealed asymmetric patterns in the bar quadrupole and the disc outskirts of the LMC in-plane velocity maps.} 
  {We aim to quantify the asymmetries detected in the LMC radial and residual tangential velocity maps, and determine whether they are generated naturally due to the LMC's interaction with the SMC.}
  {We analyse the velocity maps of different simulations from the KRATOS suite of $N$-body simulations of the LMC--SMC--MW system, proposing a new methodology to quantify the kinematic asymmetry in the bar and the outskirts of the disc. We also transform the KRATOS simulations into mock catalogues with $G$ magnitudes and \gaia observational errors, to confirm that the asymmetric signature in the LMC is not an effect of  observational uncertainties. In addition, we investigate the possibility of a classification bias in the neural network classifier of the \gaia optimal sample.}
  {In the KRATOS simulations of the LMC and SMC interaction, the dynamical effect of the SMC passages produces a displacement of the bar and asymmetries in the LMC velocity maps.
  Individual stars of the SMC do not have a substantial effect on the kinematics of the LMC. 
  By comparing the velocity maps of mock catalogues of the future \gaia data releases DR4, DR5 and \textit{GaiaNIR}, we find that the asymmetric signature in the bar quadrupole is independent of observational errors. We thereby confirm that it is a consequence of the interaction of the LMC with the SMC. We also find a classification bias in the neural network classifier, indicating that the outer disc asymmetry observed in the optimal sample is artificial.}
  {The analysis of the KRATOS simulations reveals that the interaction of the LMC with the SMC can generate asymmetric patterns in the velocity field. In the case of the \gaia DR3 LMC velocity maps we conclude that the bar quadrupole asymmetry is directly correlated with the SMC interaction, while the outer disc asymmetry is an artefact of the classifier for the optimal sample.} 


   \maketitle


\defcitealias{Jimenez-Arranz23a}{JA23}
\defcitealias{Jimenez-Arranz24a}{JA24}

\section{Introduction}
\label{sec:intro}

The LMC is the largest satellite galaxy of the Milky Way (MW) with a mass of $\sim 1.8 \times 10^{11} M_\odot$  \citep[e.g.][]{Shipp21,Watkins24} and at a distance of $\sim 49.59$ kpc \citep{Pietrzynski19}. It is a Magellanic spiral galaxy characterized by a tilted, off-centred bar and a single prominent spiral arm \citep[e.g.][]{Elmegreen1980, Zaritsky2004, GaiaLuri21, Rathore25, Jimenez-Arranz25a}. Recent studies suggest that the LMC's bar may be non-rotating \citep{Jimenez-Arranz24a, Jimenez-Arranz25b}. Together, these non‑axisymmetric structures are likely the result of the interactions of the LMC with the SMC, one of its satellite galaxies located at a distance of around $62\,$kpc from the MW \citep{Cioni00,Hilditch05,Graczyk13}. 
The LMC and SMC (hereafter MCs) are separated 20-25 kpc and, among others, a prominent feature of their interaction is the trail of stars that lies between both of them. This structure, called the Bridge, consists mostly of SMC gas and stars that have been stripped from the SMC during past tidal interactions with the LMC, (e.g. \citealp{misawa09, skowron14, carrera17, Zivick2019, GaiaLuri21}). 
Due to their proximity to us, individual stars of the MCs can be resolved (e.g. \citealp{Nidever2017, GaiaLuri21, Niederhofer2022, Jimenez-Arranz23a, Jimenez-Arranz23b}). This makes the MCs a unique opportunity to study the effects of the interaction between galaxies on their internal structures and their evolution history. 

\begin{figure}
\includegraphics[width=0.95\columnwidth]{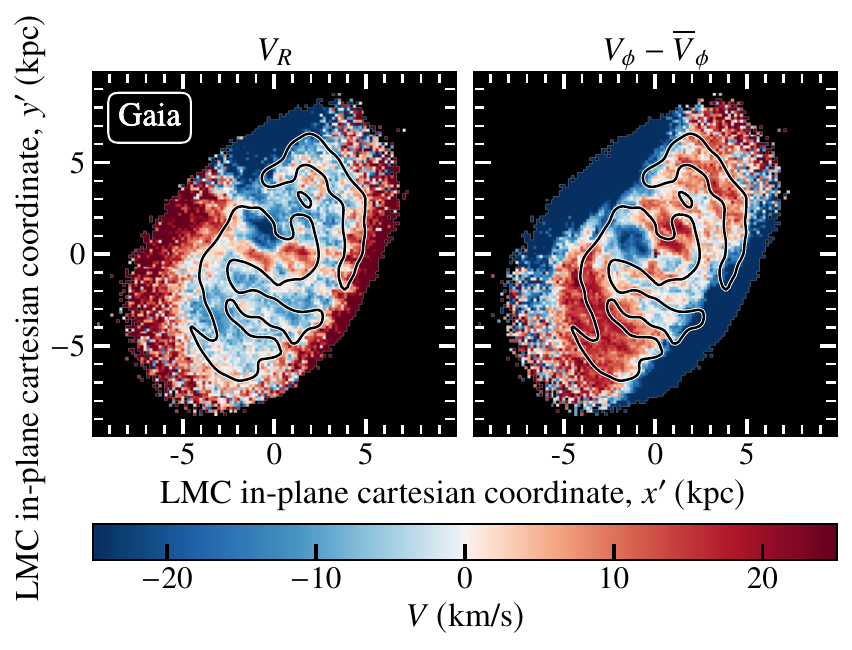}
\caption{Comparison of the radial $V_R$ (left panel) and residual tangential $\vphires$ (right panel) velocity map of the LMC optimal sample \citepalias[\gaia DR3,][]{Jimenez-Arranz23a}. The maps are shown in the LMC in-plane $(x',y')$ Cartesian coordinate system, as if seen face-on. A black contour line separates the overdensities (bar and spiral arms) from the underdensities.}
\label{fig:gaia}
\end{figure}

ESA's \gaia mission is providing astrometric, photometric and spectroscopic data for more than 1.8 billion stars, of which about $10-20$ million correspond to the MCs (\citealt{Jimenez-Arranz23a}, hereafter JA23; \citealt{Jimenez-Arranz23b}). From these measurements, we can derive 3D kinematic maps of the LMC that reveal rich complexities and substructures in the LMC disc (\citealt{GaiaLuri21}; \citealt{Choi2022}; \citealt{Cullinane2022a}; \citetalias{Jimenez-Arranz23a}; \citealt{Jimenez-Arranz25a}; \citealt{Vijayasree25}). One of these features is a largely asymmetric kinematic signature in the bar region (see Fig.~\ref{fig:gaia}). If the bar were located in the centre of mass of an unperturbed system, we would expect a perfectly symmetric quadrupole, with a positive--negative trend in each pole due to the elliptic shape of the stellar orbits in the bar. 
The kinematic maps also show strong inward and outward velocities in the outer regions of the LMC disc, particularly strong inward velocities (blue) in the upper region of the radial velocity map (left panel of Fig. \ref{fig:gaia}), which are not mirrored in the symmetric region in the lower part of the map. The unknown origin of the bar and disc asymmetries in the \gaia data motivates a pursuit to correlate asymmetric kinematic maps with irregular bars and galaxies in interaction.

Numerical simulations have been developed to model the past orbit of the LMC, investigate the dynamical effects of its interaction with other galaxies, and explore the time evolution that cannot be captured by observational data alone \citep{Besla12, garavito-camargo19,Lucchini21, vasiliev23, Sheng24, Sheng25, Brooks25}. In this work, we use the KRATOS simulations \citep[][hereafter JA24]{Jimenez-Arranz24a}, a large suite of $N$-body simulations designed with high spatial, mass and temporal resolution, aiming to study the internal kinematics of the LMC and its interactions with the SMC and the MW. 
This study aims to analyse the kinematic maps of LMC-like galaxy simulations to study how the interaction with an SMC-mass simulation can cause the asymmetries in the inner bar region as well as the outer regions of the LMC disc. We investigate whether the asymmetries observed in \citetalias{Jimenez-Arranz23a} using \gaia DR3 data can be reproduced in $N$-body simulations and therefore be a natural consequence of the LMC's evolution, particularly considering its interaction with the SMC. We also use these simulations to generate velocity maps, applying the expected uncertainities of future \gaia and \textit{GaiaNIR} releases to evaluate the effect of present \gaia uncertainties.

The article is organised as follows: In Sect.~\ref{sec:data} we describe our datasets. Sect.~\ref{sec:method} details the methods used to quantify the asymmetries in our analysis. An overview of the results is given in Sect. ~\ref{sec:results}. We add a discussion on observational uncertainties and possible sample bias in Sect.~\ref{sec:cause}, as well as a further exploration of the KRATOS simulations and future \gaia and \textit{GaiaNIR} data in Sect.~\ref{sec:future}. We summarise our work and present our conclusions in Sect.~\ref{sec:conclusions}.


\section{Data and kinematic maps}
\label{sec:data}

\subsection{LMC \gaia DR3 clean samples}
\label{subsec:gaia}
\begin{figure}
\centering
\includegraphics[width=0.98\columnwidth]{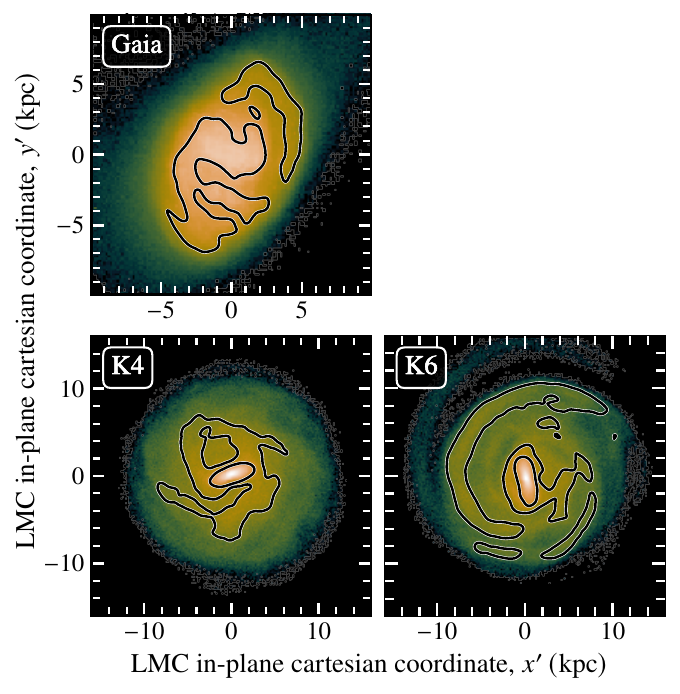}
\caption{Density maps of the LMC optimal sample using \gaia DR3 data \citepalias[top panel,][]{Jimenez-Arranz23a} and \lmc disc of the K4 and K6 simulations from the KRATOS suite \citepalias[bottom panels,][]{Jimenez-Arranz24a}. All maps are shown face-on, in the LMC in-plane $(x',y')$ Cartesian coordinate system. Black contour lines separate the overdensities (bar and spiral arms) from the underdensities. The KRATOS simulations are shown at $t = 0.105$ Ga (see Sect. ~\ref{subsec:t0} for discussion).}
\label{fig:density}
\end{figure}

\gaia DR3 \citep{GaiaVallenari23} gives us access to detailed astrometric information for a catalogue of more than $1.8$ billion stars, of which about $10-20$ million belong to the MCs (\citetalias{Jimenez-Arranz23a}; \citealt{Jimenez-Arranz23b}). 
In this work, we use the LMC optimal sample from \citetalias{Jimenez-Arranz23a}. This dataset was selected using a neural network classifier and contains $9\, 810\, 031$ candidate LMC stars. Using the coordinate transformation introduced by \citet{vandermarel01}, \citet{vandermarel02} and used in \citetalias{Jimenez-Arranz23a}, with inclination angle $i = 34^\circ$, and position angle $\theta = 220^\circ$ \citep{GaiaLuri21}, we deproject the LMC to obtain the LMC in-plane $(x',y')$ Cartesian coordinates. The top panel of Fig.~\ref{fig:density} shows the density map of the LMC optimal sample, face-on in the LMC in-plane $(x',y')$ Cartesian coordinate system.  

\subsection{KRATOS simulations}
\label{sec:KRATOSsim}

We aim to compare the \gaia DR3 LMC optimal sample with the KRATOS simulations \citepalias{Jimenez-Arranz24a}, a suite of $N$-body simulations of isolated and interacting LMC-like galaxies. KRATOS consists of 28 simulations that model the evolution of an LMC-like galaxy with varying parameters. The suite includes models in which the LMC-like galaxy is in isolation or in interaction with an SMC-mass galaxy and a MW-mass galaxy. In this work we follow the notation introduced in \citetalias{Jimenez-Arranz24a}, where the LMC-like galaxy is denoted as \lmcnospace, and the SMC- and MW-mass systems as \smc and \mwnospace, respectively. In the suite of simulations, the parameters explored are the mass of the baryonic matter or dark matter content of the \lmcnospace, the stellar Toomre parameter $Q$ of the \lmcnospace, and the mass of the \smc or the \mwnospace, if present (see Table~2 of \citetalias{Jimenez-Arranz24a} for a summary). All simulations were evolved for $4.68\,$Ga with a spatial and temporal resolution of $10\,$pc and $5000\,$yr, respectively, and a minimum mass per particle of $4 \times 10^3\,$ M$_\odot$. Details and specific characteristics of each of the 28 simulations can be found in Sect.~2 of \citetalias{Jimenez-Arranz24a}. 

The simulations analysed in this work are K4 and K6, shown in the bottom panels of Fig.~\ref{fig:density}. The K4 simulation models an isolated \lmc system, whereas the K6 simulation matches its \lmc parameters but includes the full interacting system of \lmcnospace, \smcnospace, and \mwnospace. This set of simulations was chosen for this work due to the K6 simulation's morphological similarities with the LMC, in particular its off-centred bar (bottom right panel, Fig.~\ref{fig:density}). The time stamps are defined with respect to the $t=0$ Ga established in \citetalias{Jimenez-Arranz24a} (see Sect.~\ref{subsec:t0} for further discussion).

To compare the internal kinematics of the KRATOS simulations to those of the deprojected \gaia DR3 LMC optimal sample, we determine the KRATOS simulations' \lmc in-plane position and velocity coordinates, $(x',y',v_x',v_y')$. We first centre the reference frame in the centre of mass of the \lmc baryonic matter. Then we correct for the \lmc systemic motion and align the reference frame so that the $(x',y')$-plane coincides with the \lmc disc plane by computing the disc angular momentum (see Sect. 2.3 of \citetalias{Jimenez-Arranz24a} for details). 

\subsection{Kinematic maps}
\label{sec:kinemaps}

\begin{figure}
\includegraphics[width=\columnwidth]{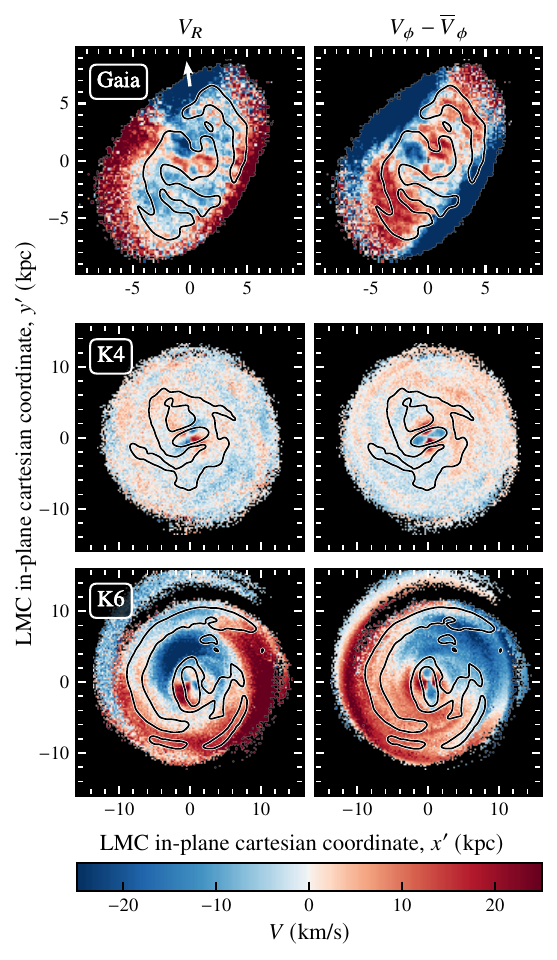}
\caption{Comparison of the radial $V_R$ (left panels) and residual tangential $\vphires$ (right panels) velocity maps between the different samples used in this work, as seen face-on. Top: LMC optimal sample \citepalias[\gaia DR3,][]{Jimenez-Arranz23a}. Centre and bottom: \lmc disc for K4 and K6 simulations \citepalias[KRATOS suite,][]{Jimenez-Arranz24a}, respectively. The white arrow marks the outskirts asymmetry and indicates the direction of the Bridge \citep{Belokurov2017, GaiaLuri21}. All maps are shown in the LMC in-plane $(x',y')$ Cartesian coordinate system. A black contour line separates the overdensities (bar and spiral arms) from the underdensities. KRATOS snapshots are shown at $t = 0.105$ Ga (see Sect. ~\ref{subsec:t0} for discussion).}
\label{fig:velocity}
\end{figure}

From the LMC and \lmc in-plane position and velocity coordinates we create galactocentric radial $V_R$ and residual tangential $V_\phi - \overline{V_\phi}$ velocity maps. The radial velocity maps show the stars' motion towards (negative) or away from (positive) the galactic centre, while the residual tangential velocity maps show if the stars are moving faster (positive) or slower (negative) than the galaxy's mean tangential velocity curve at their radial distance from the galactic centre. Figure~\ref{fig:velocity} presents the radial (left) and residual tangential (right) velocity maps for the LMC optimal sample (top panels) and the KRATOS simulations (centre and bottom panels). The binned kinematic maps are obtained by dividing the sample of stars in a square grid with $0.16 \times 0.16\,$kpc bins. Due to the different spatial extent of the \gaia data compared to the simulation, we use $125 \times 125$ bins to display $-10$ to $10\,$kpc in both $x'$ and $y'$ for the LMC optimal sample (top panels), and $200 \times 200$ bins for the simulated KRATOS discs from $-16$ to $16\,$kpc (centre and bottom panels). For each bin, we compute the median of the radial or residual tangential velocities of the enclosed stars. Contour lines (in black) based on the respective source overdensity map of the LMC (see Fig.~\ref{fig:density}), trace the bar and spiral arm(s) on the velocity maps.

There are two notable features in these kinematics maps. First, a barred galaxy presents a characteristic quadrupole pattern in the radial velocity $V_R$ and residual tangential velocity $V_\phi - \overline{V_\phi}$ maps. This signature stems from the elliptical orbits of the stars that make up the bar as seen from the centre of the bar.
This quadrupole is seen in the blue (red), negative (positive) velocities of the bins inside and near the bar contours. The quadrupole of the K4 simulation is symmetric (middle row of Fig. \ref{fig:velocity}) because no perturbation is present and the centre of mass and centre of the bar coincide.
On the other hand, for the LMC optimal sample and the K6 simulation (top and bottom rows of Fig. \ref{fig:velocity}), the centre of mass and centre of bar do not coincide (see also Fig. 9 of \citetalias{Jimenez-Arranz24a} for the oﬀ-centredness evolution of the \lmc bar), making the quadrupole signal in the radial velocity and residual tangential velocity maps asymmetric.

Second, the outer region of the LMC disc presents a high inward velocity in the upper region of the \gaia radial velocity map (top left panel of Fig. \ref{fig:velocity}), which is unmatched on the opposite side of the galaxy. A similar feature of opposite sign can be seen in the high outward radial velocity in the lower right region of the \lmc disc outskirts of the K6 simulation (bottom left panel of Fig. \ref{fig:velocity}).

\section{Quantitative analysis of the asymmetries}
\label{sec:method}

In this Section, we devise two methods to analyse the asymmetries in the LMC bar and disc outskirts. We use the kinematic maps described in Sect.~\ref{sec:kinemaps} to apply the methods first to the KRATOS simulations (Sect.~\ref{subsec:method_bar} and Sect.~\ref{subsec:method_disc}) and then to the LMC \gaia data (Sect.~\ref{subsec:method_gaia}).

\subsection{\lmc bar}
\label{subsec:method_bar}

In order to analyse the bar region of the KRATOS simulations, we bin the stars in a zoomed-in square grid, using $200 \times 200$ bins for $-4$ to $4\,$kpc, which yields bin dimensions of $0.04 \times 0.04\,$kpc. We select the bar region of the velocity maps using a contour, marking where the local density is at least 20\% higher than the Gaussian smoothed background. To quantify the asymmetry of the bar, we then compute the median radial $V_R$ and residual tangential $\vphires$ velocity of stars that are binned inside this bar contour.

For an unperturbed bar with a symmetric quadrupole, the median bar velocity would be zero because the velocities of opposite quadrants cancel out. A deviation from zero therefore indicates an asymmetry in the bar quadrupole. For the KRATOS simulations, we can analyse the time evolution of the bar quadrupole using consecutive snapshots (see Sect.~\ref{sec:resultBAR}). 

\subsection{\lmc outer disc}
\label{subsec:method_disc}

\begin{figure}
    \centering
\includegraphics[width=1\columnwidth]{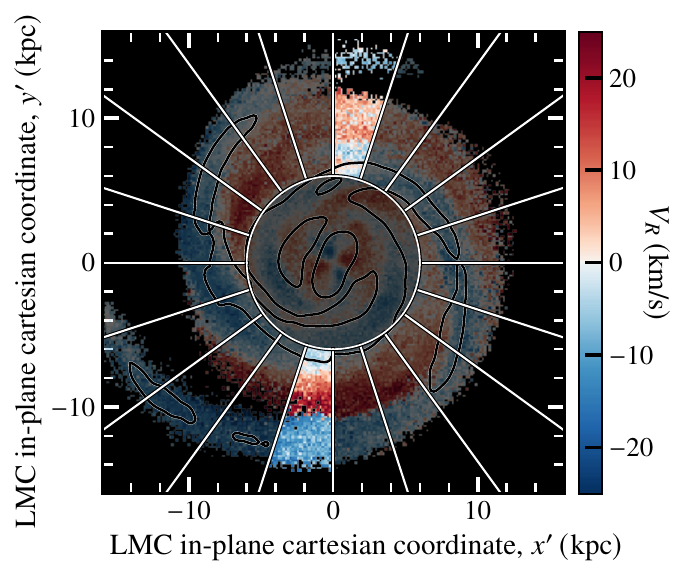}
    \caption{Visualisation of the outer disc quantitative analysis. We show the radial velocity map of a \lmc disc of the KRATOS suite, with the disc divided into 20 angular wedges of $18^\circ$ each. The first pair of opposite wedges is highlighted while the others are masked by higher opacity. The centre of the disc is further masked by a circle of radius $R = 6\,$kpc.}
    \label{fig:visualwedges}
\end{figure}

We analyse the asymmetries in the outer regions of the \lmc disc of the KRATOS simulations in the steps listed below and illustrated in Fig. ~\ref{fig:visualwedges}:
\begin{enumerate}
    \item We mask the inner region of the \lmc disc with a circle of radius $R = 6\,$kpc.
    \item We split the disc into 20 angular wedges of $18^\circ$ each. Then we compute the median value of $V_R$ and $\vphires$ in each section. 
    \item As a measure of the azimuthal symmetry of the regions of the \lmc disc, we subtract the median velocity of each wedge by its opposite wedge's median velocity (for example, the pair of wedges highlighted in Fig.~\ref{fig:visualwedges}).
    \item If the resulting subtracted median velocity deviates greatly from the expected value of zero, it indicates an asymmetry in the \lmc disc. To obtain a quantification of the maximum or minimum asymmetry of the disc per snapshot, we compute the extrema of the subtracted median velocities across the whole disc, EXT $\langle{V}_{R}\rangle$ and EXT $\langle V_\phi - \overline{V_\phi}\rangle$.
\end{enumerate}
We repeat this process for successive snapshots to analyse the time evolution of the asymmetries in Sect.~\ref{sec:resultDISC}. 

\subsection{LMC \gaia data}
\label{subsec:method_gaia}
To quantify the bar asymmetries in the LMC optimal sample from \gaia data, we fit an ellipse with an axis ratio of $q =0.46$, an angle of $\phi = 154.18^\circ$ \citep[both from][]{Choi2018b}, and a semi-major axis of 2.3 kpc \citep{Jimenez-Arranz24b}.
Then, similarly to the \lmc bar analysis (see Sect.~\ref{subsec:method_bar}) we compute the median radial, $V_R$ and residual tangential $\vphires$ velocities of the bins enclosed in this ellipse. 
To analyse the asymmetries in the outer regions of the LMC optimal sample, we follow the same steps as outlined in Sect.~\ref{subsec:method_disc}, but masking the inner $4\,$kpc of the disc, instead of the inner $6\,$kpc. This analysis provides two extrema for the \gaia outer disc, one for the median radial velocity, EXT $\langle{V}_{R}\rangle$, and one for the median residual tangential velocity, EXT $\langle V_\phi - \overline{V_\phi}\rangle$. 
Since we do not have any time evolution in the \gaia data, we recover single values for the median velocities of the LMC bar, and for the extrema of the outer disc median velocities.

\newcommand{\vrmean}{$\langle V_R \rangle$ }
\newcommand{\vphimean}{$\langle V_\phi- \overline{V_\phi} \rangle$ }

\section{Results}
\label{sec:results}

\subsection{Asymmetry in bar kinematics}
\label{sec:resultBAR}

\begin{figure}
\centering
\includegraphics[width=\columnwidth]{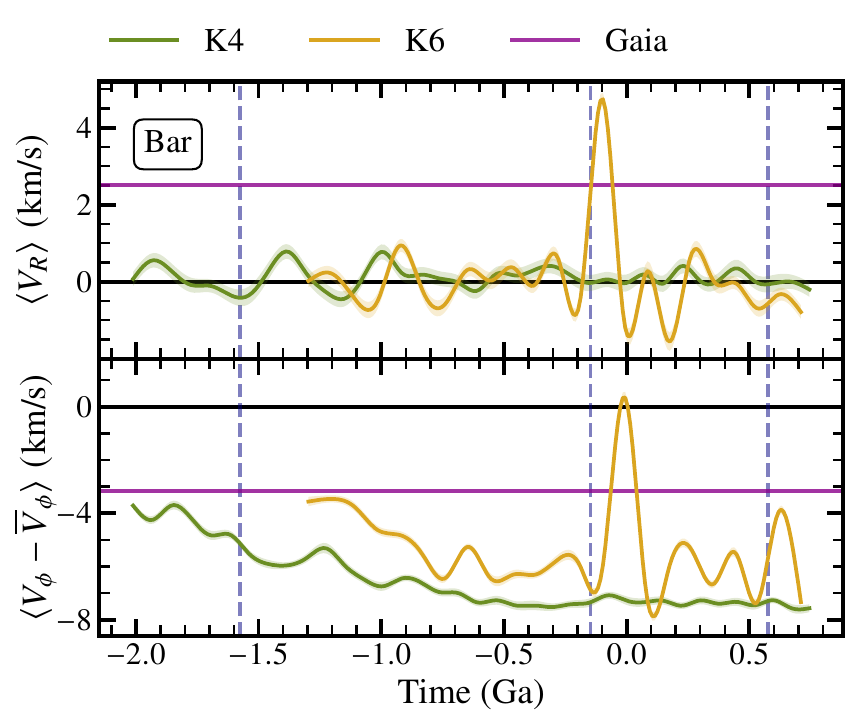}
\caption{Mean radial velocity $\langle V_R \rangle$ (top panel) and mean residual tangential velocity $\langle \vphires \rangle$ (bottom panel) asymmetry inside the \lmc bar for the K4 (green lines) and K6 (yellow lines) simulations, as a function of time. The \gaia asymmetry is shown with horizontal purple lines. The horizontal black lines mark zero net velocity. The vertical dashed blue lines represent the pericentric passages between \lmcnospace--\smcnospace. We only show data points after the bar formation (relative $m=2$ Fourier amplitude $\Sigma_2/\Sigma_0 > 0.2$).}
\label{fig:barevol}
\end{figure}

To analyse how the asymmetry of the \lmc bar changes with time, we show the evolution of the K4 and K6 simulations' median radial \vrmean and residual tangential \vphimean velocity of the bar in Fig.~\ref{fig:barevol}. The quadrupole of the K4 simulated bar is symmetric with only small deviations from zero in \vrmean (top panel, green line) after bar formation at approximately $t\sim-2\,$Ga. Similarly, the \vphimean (bottom panel, green line) does not show abrupt variations, but gradually becomes more negative. This is because the bar lengthens over time (see for example Fig. 8 of \citetalias{Jimenez-Arranz24a}), causing the bar contour to capture more of the negative residual tangential component.

In the interacting K6 simulation, both \vrmean and \vphimean (top and bottom panel, yellow line) vary significantly after bar formation at approximately $t=-1.34\,$Ga. The bar formation of the K6 simulation occurs later than that of K4, namely after the first pericentric passage of the \lmcnospace--\smcnospace. After this interaction, both the median radial and median residual tangential velocity deviate from zero. Particularly after the second pericentric passage between the \lmcnospace--\smc at $t\sim-0.15\,$Ga, there is a maximum in the median radial velocity of the bar, immediately followed by a maximum in the median residual tangential velocity. For comparison, we mark the asymmetry of the \gaia data with horizontal, purple lines in both panels. For both the radial and the residual tangential velocity, the \gaia bar asymmetry line passes through the peaks of the K6 simulation asymmetry, shortly after the second pericentric passage.

\subsection{Asymmetry in outer disc kinematics}
\label{sec:resultDISC}
\begin{figure}
\centering
\includegraphics[width=\columnwidth]{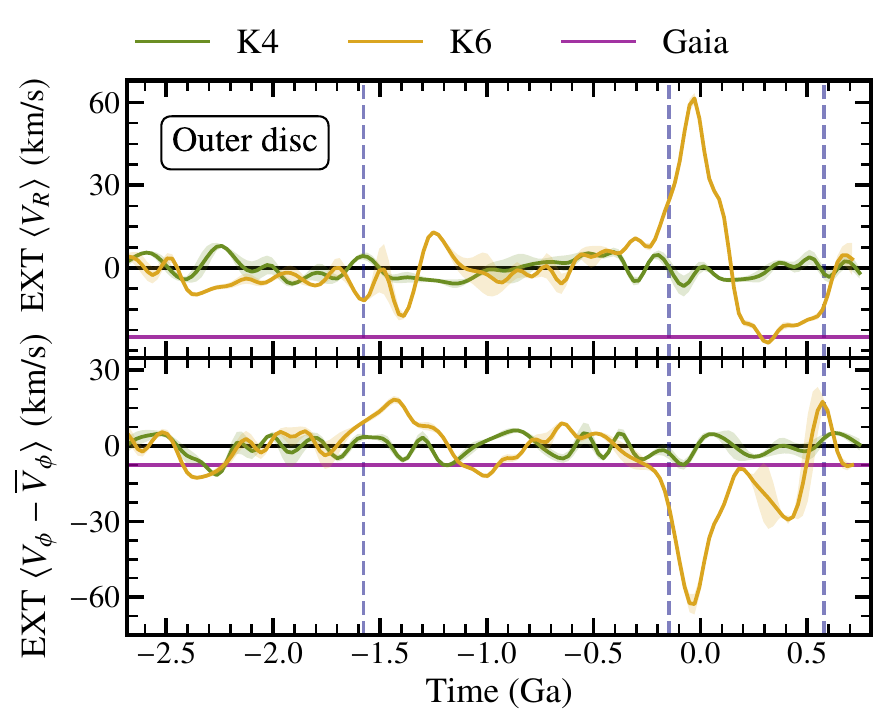}
\caption{Same as Fig.~\ref{fig:barevol} but for the outer disc asymmetry evolution of the \lmcnospace, represented by the extrema in the disc velocities for each snapshot from the K4 (green lines) and K6 (yellow lines) simulations.}
\label{fig:discevol}
\end{figure}

To analyse the changes in the kinematics of the \lmc outer disc, we apply the method described in Sect.~\ref{subsec:method_disc} to the full temporal evolution of the K4 and K6 simulations. Computing the extrema in the kinematic map velocities for different timestamps provides us with the evolution of the disc outskirts' asymmetry over time, shown in Fig.~\ref{fig:discevol}.
Similarly to the time evolution of the bar in Sect.~\ref{sec:resultBAR}, the unperturbed K4 simulation shows no significant variation in the extrema of $\langle{V}_{R}\rangle$ or $\langle V_\phi - \overline{V_\phi}\rangle$ (top and bottom panels, green lines), indicating symmetry in the disc outskirts. In contrast, the interacting K6 simulation has a maximum in the extrema of the mean radial velocities (upper panel, yellow line) after the second pericentric passage of the \smc at $t\sim-0.15\,$Ga. At this time, the \lmc disc outskirts have a large asymmetric feature of positive radial velocity, which is unmatched on the opposite side of the disc. The same time stamp is also characterized by a minimum in the mean residual velocities (bottom panel, yellow line), which signifies a strong asymmetry skewing to negative residual tangential velocity of the disc. Another local extremum of lower amplitude can be seen after the first pericentric passage of the \smc around $t\sim-1.58\,$Ga, in both the radial and residual tangential velocities. Again, we also show the asymmetry of the \gaia data by marking horizontal purple lines for the extrema of the median radial and median residual tangential velocities of the disc outskirts. Here we see that the \gaia asymmetry line passes through the minimum of the K6 radial velocity extrema, but it is close to zero in the residual tangential velocity extrema. 

\subsection{Re-evaluating t = 0}
\label{subsec:t0}
The initial conditions and orbital parameters of \lmcnospace, \smc and \mw in KRATOS were based on the work by \citet{Lucchini21} and the ``present time'' $t = 0$ Ga was determined to match the number of \lmcnospace--\smc pericentric passages and the morphology of the \lmc disc, rather than the \lmc closest approach to the \mw (see Sect. 3 of \citetalias{Jimenez-Arranz24a} for more details). While the ``present time'' considered in KRATOS qualitatively matches the morphology of the present-day LMC disc, in this work we aim to redefine the timestamp $t = 0$ Ga in the KRATOS simulation to match more closely with the kinematic maps of the \gaia DR3 LMC optimal sample. 

Due to the complexity of the asymmetric features on a perturbed disc, we cannot simply define the new $t = 0$ based on the intersection of the \gaia asymmetry line (purple line) with the asymmetry evolution plots of the K6 simulation in Figs.~\ref{fig:barevol} and ~\ref{fig:discevol}. Around these intersection points in the K6 simulation, we mainly find snapshots where the bar has been completely destroyed, or the disc outskirts have bimodal velocities due to the recent passage of the \smcnospace. To best match the complex asymmetries of the kinematic maps, we instead select a broad range of timestamps around the second pericentric passage and visually inspect the kinematic maps. 

We find a closer representation of the LMC optimal sample asymmetry at $t = 0.105$ Ga, shown in the bottom row of Fig.~\ref{fig:velocity}. Both the LMC optimal sample (top row) and the K6 simulation (bottom row) have an off-centred bar and one broken spiral arm. The velocity maps show similar trends but with opposite signs: in the LMC optimal sample, the bar's quadrupole has strong asymmetry in the negative radial velocity, whereas the K6 snapshot has a similar asymmetric feature in the positive radial velocity. There are similar patterns of negative residual tangential velocity in both bars, including some asymmetry. In the disc outskirts, we see negative radial velocities in the top left of both maps, whereas the rest of the outskirts are dominated by positive radial velocities. We are unable to match the outskirts in the residual tangential velocity with the strong negative velocities along opposite sides of the disc. However, this feature is likely a result of the galaxy's morphology, because we compute the residual tangential velocity by subtracting radially binned mean tangential velocities, despite the LMC's elongated disc.
For future studies using the KRATOS simulations, we recommend that the snapshot at time $t = 0.105$ Ga \footnote{The snapshot at time $t = 0.105$ Ga is also labelled $a = 0.857$ in the KRATOS suite.} from the K6 simulation be used as the ``present time'', because it not only represents the morphology of the LMC but also resembles its velocity maps more closely.


\section{Cause of asymmetries}
\label{sec:cause}
In this section, we explore several possible causes of the asymmetries seen in the kinematic maps of the \gaia LMC optimal sample \citepalias{Jimenez-Arranz23a} and characterised in this work. First, in Sect.~\ref{subsec:interact}, we present our study of LMC--SMC interactions using the KRATOS suite \citepalias{Jimenez-Arranz24a} to explain the asymmetries described in Sect.~\ref{sec:results}.
Then we discuss the possibility of a bias in the LMC optimal sample stemming from the neural network classifier in Sect.~\ref{subsec:classifier}, and address the potential effect of the observational errors of \gaia in Sect.~\ref{subsec:mock}. Lastly, we repeat the asymmetry analysis for a larger sample of KRATOS simulations in Sect.~\ref{subsec:robust} to verify the consistency of our results.

\subsection{LMC-SMC interaction}
\label{subsec:interact}
From the K4 simulation in Figs.~\ref{fig:barevol} and~\ref{fig:discevol} (green lines) we can see that an isolated \lmc does not generate a significant asymmetry in the bar, or in the outer disc. Without an external perturbation such as the \lmc -- \smc interaction, the median radial and median residual tangential velocities of the \lmc do not show significant variation over time. In the K6 simulation, on the other hand, there are extrema in the velocity maps of both the bar and the outer disc (yellow lines). We can furthermore see that the extrema are correlated in time with the pericentric passage of the \smcnospace. Therefore, we can conclude from the KRATOS simulations that the asymmetries observed in the LMC disc of \gaia data may arise naturally due to the interaction of the LMC with the SMC.

\begin{figure}
        \centering
    \includegraphics[width=\columnwidth]{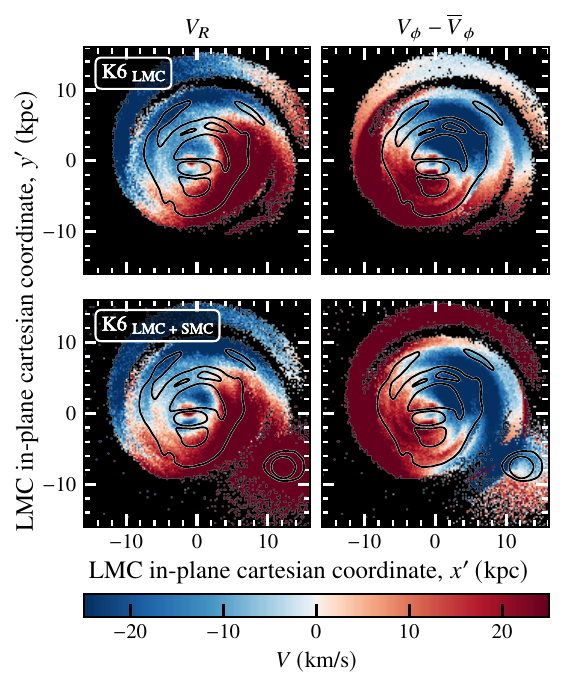}
    \caption{Comparison of the radial $V_R$ (left panels) and residual tangential $\vphires$ (right panels) velocity maps between the \lmc and the \lmc -- \smc of the K6 simulation right after the second pericentric passage, at $t = 0$ Ga \citepalias[KRATOS suite,][]{Jimenez-Arranz24a}. We show the same snapshot of the K6 simulation for two cases: \lmc disc only (top panels), and \lmc disc and \smc particles (bottom panels). 
    All maps are seen face-on, in the LMC in-plane $(x',y')$ Cartesian coordinate system. A black contour line separates the overdensities (bar and spiral arms) from the underdensities.}
    \label{fig:smc}
\end{figure}

Additionally, since the inward motion in the radial velocity map of the LMC optimal sample is located in the direction of the Bridge, the authors in \citetalias{Jimenez-Arranz23a} hypothesised that it could be a signature of SMC stars accreting into the LMC. Using the K6 simulation, we evaluate whether the \smc stars in the \lmc kinematic maps could gnerate the asymmetries seen in \gaia data, by presenting velocity maps that include the stellar particles of the \smcnospace. We show in Fig.~\ref{fig:smc} that, while the \smc stars are crossing the disc of the \lmcnospace, their presence alone does not change the kinematic maps significantly. Instead, the \smcnospace 's pericentric passage induces an overall dynamical effect on the kinematic maps of the \lmcnospace. However, the sign and amplitude of the changes in the velocity maps depend on the orbit of the \smc around the \lmcnospace, and in the KRATOS simulations we do not have the true orbits to qualitatively reproduce the exact signature we observe in the \gaia data. 

\subsection{Bias in the LMC optimal sample classifier}
\label{subsec:classifier}
\begin{figure}
        \centering
    \includegraphics[width=\columnwidth]{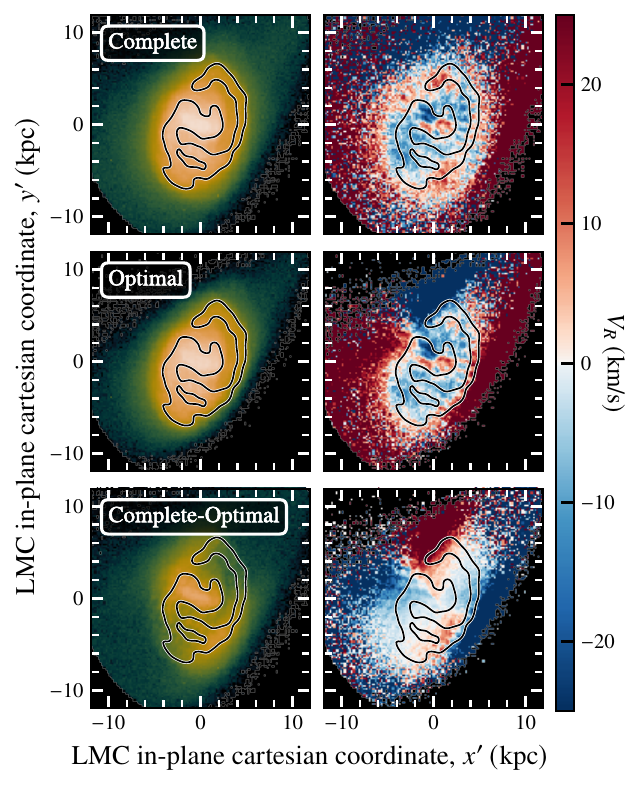}
    \caption{Comparison of the density maps (left panels) and radial velocity maps (right panels) according to the neural network classification in \citetalias{Jimenez-Arranz23a} of the complete sample (top row), optimal sample (middle row), and the difference between complete and optimal sample (bottom row).
    All maps are seen face-on, in the LMC in-plane $(x',y')$ Cartesian coordinate system. A black contour line separates the LMC overdensities (bar and spiral arms) from the underdensities.}
    \label{fig:NNbias}
\end{figure}

A classification bias in the neural network classifier of the LMC sample from \citetalias{Jimenez-Arranz23a} could affect the generated velocity maps, because a misclassification of LMC stars in some regions could result in a distribution of velocities that is biased with respect to the distribution of the true LMC population.
To investigate this possible bias, we compare the density and radial velocity maps of different LMC samples obtained from the neural network classifier in Fig.~\ref{fig:NNbias}. We show the complete sample (top row), as defined by \citetalias{Jimenez-Arranz23a} with a probability cut of $P_\text{cut}=0.01$, and the optimal sample (middle row), with a cut of $P_\text{cut}=0.52$, where stars with $P>P_\text{cut}$ are considered part of the LMC. The bottom row shows the difference between the two samples. 

The radial velocity maps (right column) show that the asymmetry in the disc outskirts becomes much more pronounced in the optimal sample (middle row), which is a clear indication of a possible bias, since there is no physical reason for this change between the two samples. To highlight this, in the bottom right panel we show the difference between the complete and optimal sample's velocity maps, which has large structures of positive radial velocity in the upper left corner and negative radial velocities along the sides. These features are most likely artefacts of the classifier and would not be present in an unbiased sample. Even though the LMC--SMC interaction can explain asymmetries in the disc outskirts (see Sect.~\ref{subsec:interact}), we can therefore conclude that the asymmetry in the disc outskirts of the \gaia optimal sample radial velocity map is a consequence of the sample selection. 

The misclassification can likely be attributed to limitations of the neural network’s training sample. As shown in Figure 2 of \citetalias{Jimenez-Arranz23a}, The proper motion distribution of the base \gaia sample in right ascension and declination is broader than that of the training set, which is centred on the systemic motion of the LMC. As a result, the classifier performs best for stars near the centre of the LMC, while objects in the outskirts are more likely to be misclassified due to the difference in proper motion from the centre to the outskirts of the LMC. Since a purer sample more closely resembles the training sample, it also inherits its limitations. We therefore recommend the use of the complete sample in future studies of the outer disc.

\subsection{\gaia observational errors and selection function}
\label{subsec:mock}

To verify that the asymmetries in the velocity maps are not a result of observational uncertainties, we apply \gaia DR3 observational errors to the KRATOS simulations. To this end, we project the \lmc to the position and orientation of the LMC in the heliocentric reference frame, using centre coordinates ($\alpha_0, \delta_0$) = ($81.28^\circ, -69.78^\circ$) \citep{vandermarel01}, inclination angle $i = 34^\circ$, and position angle $\theta = 220^\circ$ \citep{GaiaLuri21}. The effect of a different inclination or position angle on the kinematic maps has been studied in \cite{Jimenez-Arranz23a}, where no resulting systematic asymmetry was observed. 
From the projected KRATOS simulations, we create a mock catalogue of \gaia observable parameters\footnote{The mock catalogue tool is open-access on GitHub: \url{https://github.com/mschoelch24/MockCatalogue}.}: positions ($\alpha, \delta$), parallax ($\varpi$), and proper motions ($\mu_{\alpha*}, \mu_{\delta}$). 

We use a 3D extinction model \citep{Marshall06, Lallement22}, which is also applied in the Gaia Object Generator (GOG, \citealp{gog}). To assign \gaia photometry to the simulated particles, we assume red clump (RC) stellar parameters, since RC stars are dominant in the LMC \citep{luri20}. From the absolute magnitude $M_k = -1.61$ and intrinsic colour $(J-K)_0 = 0.55$ of RC stars \citep{Ruiz-Dern18,Straizys09}, we can compute $G$ magnitudes for each particle. Then we use the \gaia DR3 error model of \textsc{PyGaia} \citep{Brown12-24} to assign uncertainties in position ($\sigma_{\alpha*}, \sigma_{\delta}$), parallax ($\sigma_{\varpi}$), and proper motion ($\sigma_{\mu_{\alpha*}}, \sigma_{\mu_{\delta}}$)\footnote{The parallax uncertainties and scaling factors applied to compute position and proper motion uncertainties are sourced from the \href{https://www.cosmos.esa.int/web/gaia/science-performance}{\gaia science performance} webpage.}. 

\begin{figure}
        \centering
    \includegraphics[width=0.95\columnwidth]{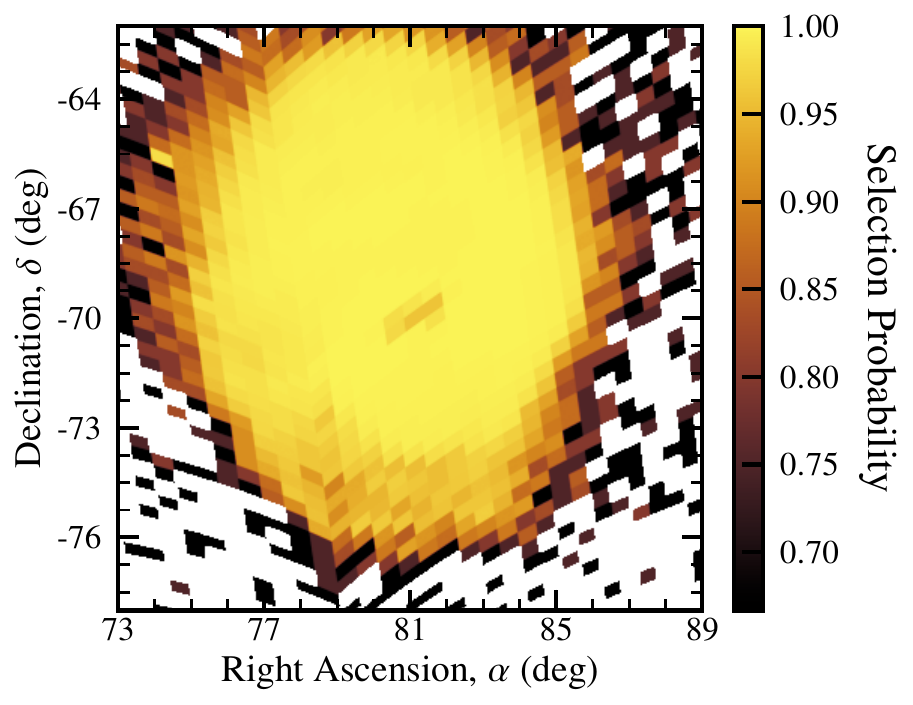}
    \caption{Sub-selection function from \textsc{GaiaUnlimited} for RC stellar parameters ($G_\text{BP}-G_\text{RP}$ colour from -0.1 to 0.4 and magnitude $G = 18-20 $ mag), for the LMC region with centre coordinates ($\alpha_0, \delta_0$) = ($81.28^\circ, -69.78^\circ$) \citep{vandermarel01}.}
    \label{fig:subsel}
\end{figure}

In order to evaluate the completeness of \gaia in the LMC region, we apply the \gaia selection function from \cite{Castro-Ginard23}, which estimates the probability that an input source is included in \gaia DR3.
Using \textsc{GaiaUnlimited}\footnote{The python package \href{https://gaiaunlimited.readthedocs.io/en/latest/index.html} {\textsc{GaiaUnlimited}} allows users to query or build \gaia selection functions.} \citep{Cantat-Gaudin23, Castro-Ginard23}, we create a sub-selection function, binning by colour $G_{\text{BP}} - G_{\text{RP}}$ and magnitude $G$ of RC stars. We show the result of this selection function per healpix region in Fig.~\ref{fig:subsel}. In the LMC disc, \gaia has a high completeness with a selection probability of close to 1.0, even in the crowded bar region. However, since we are limiting this sub-selection function to RC stellar parameters, the outskirts and surroundings of the LMC are undersampled. Therefore, the sub-selection function cannot be directly applied to the KRATOS simulations, which do not match the LMC's morphology and spatial extent. Instead, we apply the general \gaia survey selection function to the mock catalogue. This function is homogeneous in the LMC region with high completion everywhere except in the crowded bar.

\begin{figure}
        \centering
    \includegraphics[width=0.98\columnwidth]{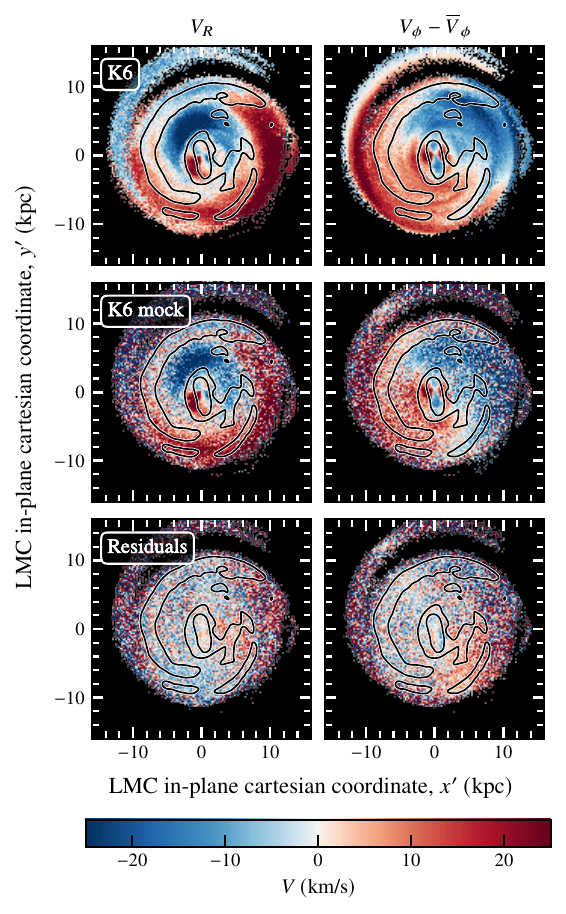}
    \caption{Comparison of the radial $V_R$ (left panels) and residual tangential $\vphires$ (right panels) velocity maps of the K6 simulation (top row), the K6 mock catalogue with \gaia DR3 observational errors and the selection function (middle row), and the residuals from the difference between the two maps (bottom row). All maps are shown in the LMC in-plane $(x',y')$ Cartesian coordinate system. Black contour lines separate the overdensities (bar and spiral arms) from the underdensities.}
    \label{fig:residuals_sf}
\end{figure}

We then convert our \gaia mock catalogue back to the LMC in-plane coordinate frame, shown in Fig.~\ref{fig:residuals_sf}. We compare the original K6 simulation (top row) with the mock catalogue (middle row), and their residuals (bottom row). 
We find that the effect of crowding in the bar region of the LMC removes less than 1\% of the stars in our sample, and the asymmetric signature in the bar is still visible (middle row of Fig.~\ref{fig:residuals_sf}). 
The observational uncertainties and selection function of \gaia data are therefore not sufficient to explain the asymmetric signal in the bar region. In the disc outskirts, the general \gaia selection function has high completeness, but the astrometric uncertainties have a larger impact.

\subsection{Consistency of results among the KRATOS suite} 
\label{subsec:robust}
\begin{figure*}
  \includegraphics[width=\textwidth]{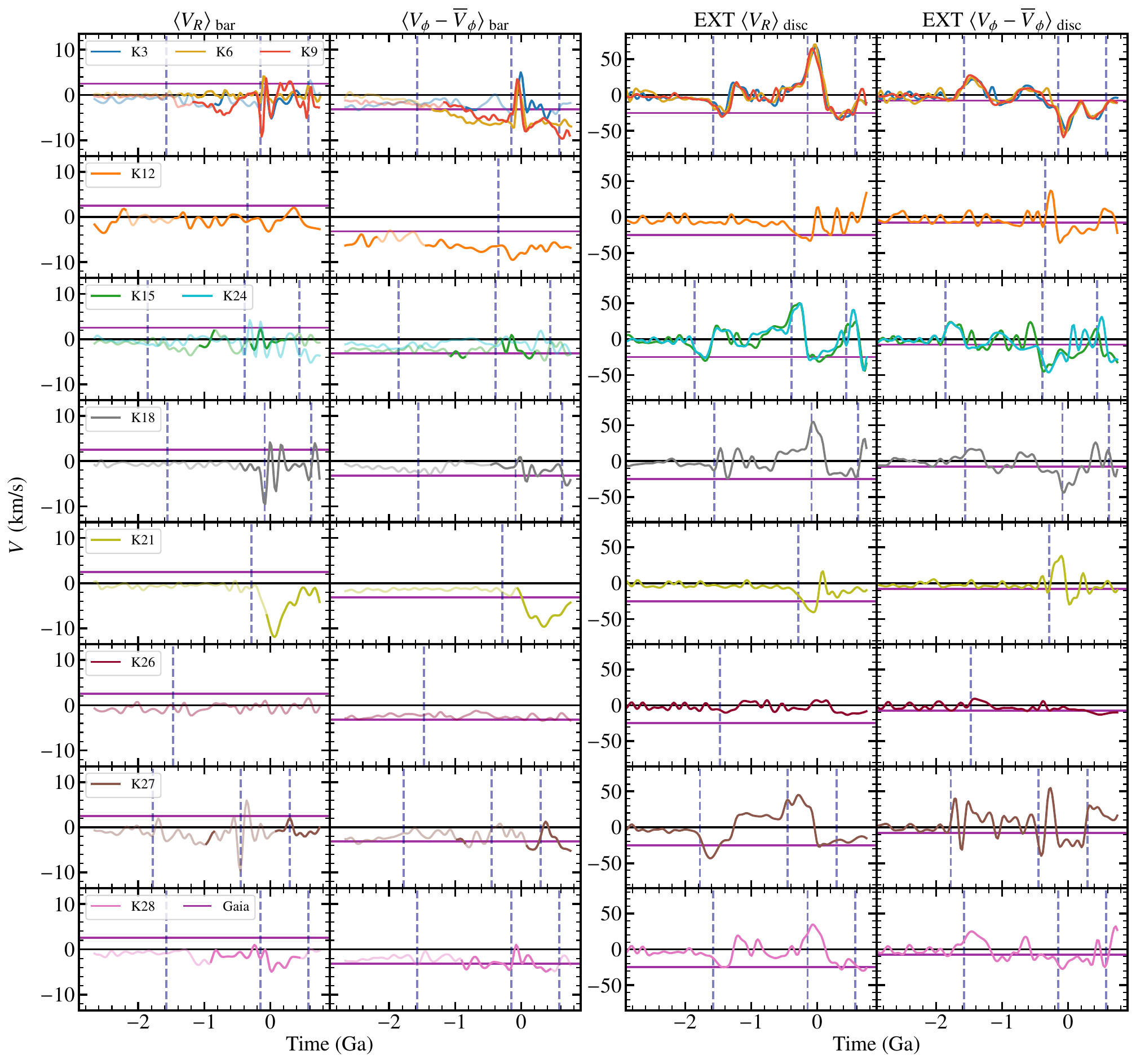}
  \caption{Bar asymmetry evolution (left two columns) and outer disc asymmetry evolution (right two columns), shown with cubic spline fits for all KRATOS simulations that include the full \mwnospace--\lmcnospace--\smc interaction. Models are grouped together if they share pericentric passages between \lmcnospace--\smc (vertical dashed blue lines) at the same timestamps. 
  The bar asymmetry evolution is measured in mean radial velocity $\langle V_R \rangle$ (left-most panels) and mean residual tangential velocity $\langle \vphires \rangle$ (centre-left panels). Data is shown only after the first 1.25 Ga, when the disc is relaxed. It is presented with a solid line when a bar is formed (relative $m=2$ Fourier amplitude $\Sigma_2/\Sigma_0 > 0.2$), and with a transparent line when there is no bar.
  The outer disc asymmetry evolution is represented by the extrema in the mean radial velocities EXT$\langle V_R \rangle$ (centre-right panels) and extrema in the mean residual tangential velocities EXT$\langle \vphires \rangle$ (right-most panels). 
  In all panels, the horizontal black lines mark zero asymmetry and the \gaia asymmetries are shown with horizontal purple lines.}
    \label{fig:allKRATOS}
\end{figure*}

In order to test the ubiquity of asymmetry in interacting simulations, we analyse the additional interacting models of the KRATOS suite. In Fig.~\ref{fig:allKRATOS}, we show our analysis of the asymmetry evolution in the bar (left two columns) and in the disc outskirts (right two columns) and note the following features:
\begin{itemize}
    \item K3, K6, and K9 (first row) have the same masses and therefore the same orbits and pericentric passages. Differences between the models affect only the internal structure of the \lmc due to their different stellar Toomre parameter $Q$. Consequently, there are minimal differences in the outer disc asymmetry among the models, but K6 (yellow line) shows some variation even before the first pericentric passage, since it has the lowest Toomre parameter, $Q=1.0$. All models show large variation correlated with the pericentric passages, particularly the second. The same is true for the bar asymmetry evolution, where all models form bars only after the first pericentric passage, and have larger variation after the second and third.
    \item K12 (second row) and K21 (fifth row) have lower \lmc dark matter halo masses, and therefore have only one late pericentric passage. Additionally, these pericentric passages happen at larger distances of about 13-14 kpc from the \lmcnospace, compared to the fiducial (K3) pericentric passages at distances of about 12 and 5 kpc. This leads to very small asymmetries in the outer disc for both K12 and K21. Of the two, K21 (lime green) has larger variation in bar asymmetry after the pericentric passage because the model also has a lower \lmc disc mass.
    \item K15 and K24 (third row) have the same \lmc dark matter halo, which has a higher mass than the fiducial and therefore causes their identical pericentric passages to happen earlier than in the fiducial simulations. Even though the K15 and K24 have different \lmc disc masses, their outer disc asymmetry evolves very similarly. K15 forms a bar despite the large mass of its dark matter halo, and then has some bar asymmetry after the second pericentric passage. K24 does not form a bar due to its lower disc mass.
    \item K18 (fourth row) has very similar pericentric passages to K3, K6, and K9, because its only difference from the fiducial is a smaller \lmc disc mass. The bar is formed later than in the fiducial model, but the asymmetry variations are similarly more prominent after the second and third pericentric passages. 
    \item K26 (sixth row) has a reduced \smc mass and therefore only one pericentric passage, which happens at a large distance of around 17 kpc and therefore causes the least variation in the outer disc asymmetry evolution. In this model, the \lmc never forms a bar. 
    \item K27 (seventh row) has a lower \mw mass and therefore has the closest \lmcnospace--\smc pericentric passages, at distances around 6.5 and 1 kpc, causing high asymmetry already after the first pericentric passage. The second pericentric passage destroys the bar shortly after its formation, but it is reformed before the third pericentric passage.
    \item K28 (last row) has the same masses as K3, K6, and K9, but a point-like \mwnospace. This changes the orbits so that the pericentres are at larger distances, around 13 and 12 kpc. As a result, the asymmetries have lower amplitudes compared to K3, K6, and K9.
\end{itemize}    

Across all simulations, there are two consistent trends: asymmetries in the velocity maps are correlated in time with pericentric passages, and closer pericentric passages cause larger variations in the asymmetry.

\section{Improvement of LMC internal kinematics}
\label{sec:future}
In this section, we discuss how the asymmetry signal changes with improvements in the observational data. We apply the tool presented in Sect.~\ref{subsec:mock} to the K6 simulation to obtain mock catalogues representative of \gaia DR4 and DR5 (Sect.~\ref{subsec:dr4dr5}), and the proposed \textit{GaiaNIR} (Sect.~\ref{subsec:nir}). 

\subsection{\gaia DR4 and DR5}
\label{subsec:dr4dr5}
Using the mock catalogues, we can make further predictions of the ability of future \gaia data releases to capture the signatures of asymmetry in the LMC disc. We use the parallax uncertainty relation and apply the scaling factors listed in Table~\ref{table:factors} from \textsc{PyGaia} \citep{Brown12-24} to compute the uncertainties in positions and proper motions of future \gaia data releases. 
\begin{table}
\caption{\gaia uncertainty scaling factors for the different \gaia data releases \citep{Brown12-24}.}
\centering
\begin{tabular}{lcccccc}
\toprule
\toprule
\gaia & \textbf{$\varpi$} & \textbf{$\alpha*$} & \textbf{$\delta$} & \textbf{$\mu_{\alpha*}$} & \textbf{$\mu_{\delta}$} \\
\midrule
DR3   & 1.000 & 0.800 & 0.700 & 1.030 & 0.890\\
DR4 &  0.749 & 0.800 & 0.700 & 0.580 & 0.500\\
DR5 & 0.527 & 0.800 & 0.700 & 0.290 & 0.250\\
\bottomrule
\end{tabular}

\tablefoot{Unitless scaling factors for parallax $\varpi$, right ascension $\alpha$, declination $\delta$, proper motion in right ascension $\mu_\alpha$, and proper motion in declination $\mu_\delta$. Uncertainties in the positions and proper motions (in mas and mas/yr) are a function of the uncertainty in parallax (in mas), e.g. $\sigma_{\alpha*} = 0.800 \cdot \sigma_\varpi$.}
\label{table:factors}
\end{table}

\begin{figure}
        \centering
    \includegraphics[width=0.9\columnwidth]{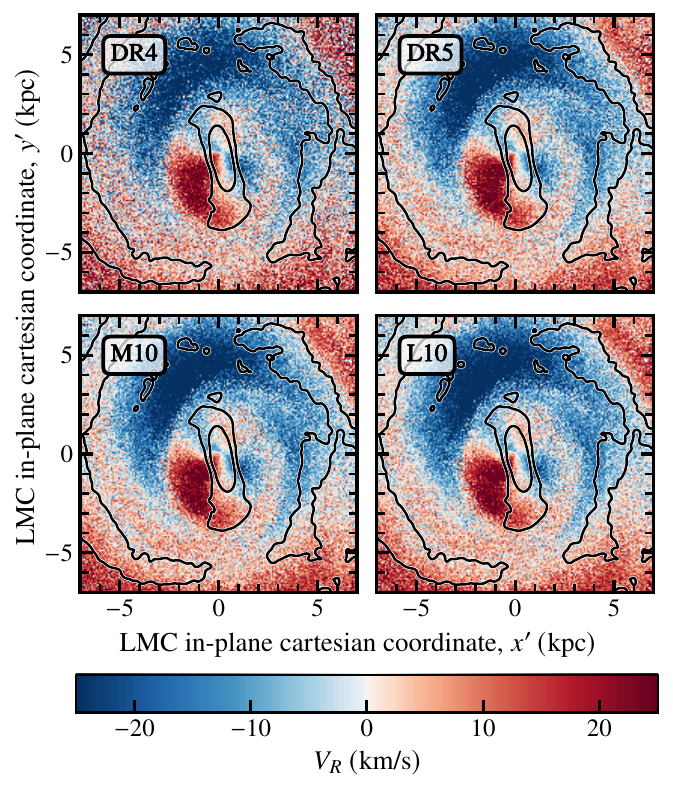} 
    \caption{Comparison of the radial velocity maps of mock catalogues from the K6 simulation with observational errors of \gaia (DR4 and DR5, top row) and \textit{GaiaNIR} (M10 and L10, bottom row). All maps are shown in the LMC in-plane $(x',y')$ Cartesian coordinate system and zoomed-in to the center coordinates to show the bar signature. Black contour lines separate the overdensities (bar and spiral arms) from the underdensities.}
    \label{fig:residuals}
\end{figure}

Although the uncertainties of \gaia DR4 and DR5 will be significantly reduced, the velocity maps of the mock catalogues do not show large differences (top rows of Fig.~\ref{fig:residuals}). In particular, the bar signature in the centre remains virtually unchanged, although sharper than in DR3 (Fig.~\ref{fig:residuals_sf}). We can conclude that the asymmetric features in the velocity maps are not simply a result of the large uncertainties in the \gaia DR3 data, because lower uncertainties reveal the same asymmetries.
However, we expect to resolve more details about the disc outskirts in future \gaia data releases.

\subsection{\textit{GaiaNIR}}
\label{subsec:nir}
While \gaia observations are limited to the visible spectra, a planned follow-up astrometric mission will extend the wavelength range into the near-infrared (NIR, \citealp{Hobbs21}; Hobbs et al. (in prep.)). The \textit{GaiaNIR} mission is aimed at exploring dust-obscured regions of the MW, including the bulge, spiral arms, and other star-forming regions. The all-sky observations will naturally include a large sample of LMC stars, particularly dimmer stars due to the depth of anticipated \textit{GaiaNIR} observations. Using the predicted uncertainties of the \textit{GaiaNIR} mission from Hobbs et al. (in prep.), we can evaluate the applications of \textit{GaiaNIR} data to our science goals and discuss the resulting advances in the study of LMC kinematics.

\begin{figure}
        \centering
    \includegraphics[width=\columnwidth]{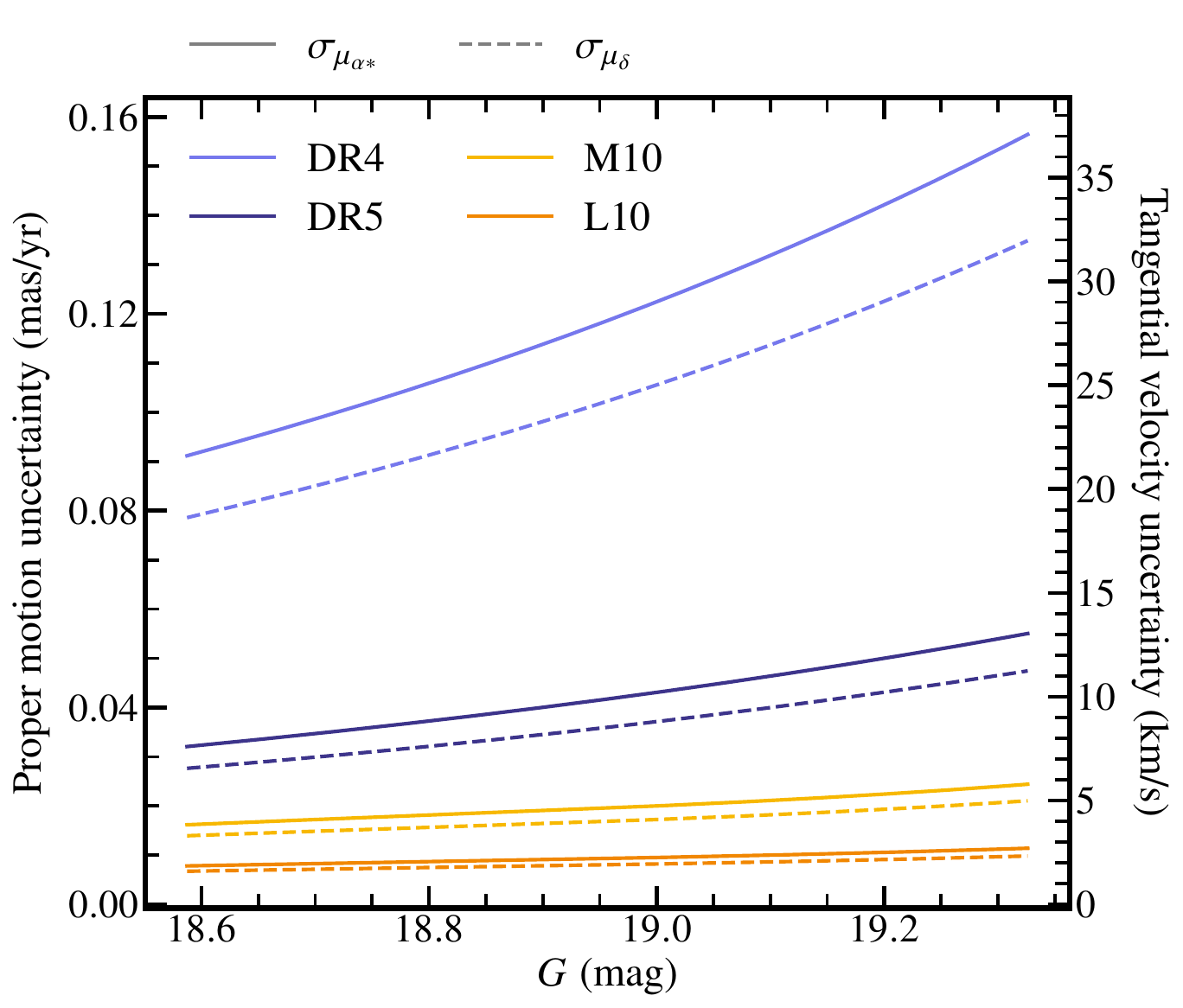}
    \caption{Proper motion uncertainty and tangential velocity uncertainty as a function of $G$ magnitude, for mock catalogues applying errors of different releases of \gaia (DR4 and DR5) and \textit{GaiaNIR} (M10 and L10). We show the uncertainties in proper motion in right ascension $\sigma_{\mu_{\alpha*}}$ (solid lines) and in declination $\sigma_{\mu_{\delta}}$ (dashed lines).}
    \label{fig:NIRerrors}
\end{figure}

We compute \textit{GaiaNIR} uncertainties for two proposed mission sizes, medium and large, each for a 10-year mission duration.
We run the uncertainty computation by Hobbs et al. (in prep.) with a K5III stellar spectrum to simulate RC stars. In Fig.~\ref{fig:NIRerrors} we show the uncertainties in the proper motions of the mock catalogues derived in Sect.~\ref{subsec:mock}. The uncertainties decrease significantly from \gaia DR4 and DR5 to the \textit{GaiaNIR} missions, with an additional improvement from the medium (M10, yellow) to the large \textit{GaiaNIR} mission (L10, orange). The velocities in our LMC radial and residual tangential velocity maps are derived from the proper motions, and the mock catalogues and residuals of the $V_R$ maps of the \textit{GaiaNIR} M10 and L10 missions are shown in the bottom rows of Fig.~\ref{fig:residuals}. Since we evaluate a velocity range from -25 to 25 km/s, and the changes between the missions are on the order of 2 km/s, we do not see visible differences in the binned velocity maps. 

Furthermore, in applying predictions of \textit{GaiaNIR} uncertainties to the mock catalogue, we are faced with the limitation of using only RC stars as the stellar tracer of the LMC. For RC stars, the \textit{GaiaNIR} uncertainties do not significantly improve the velocity maps with respect to \gaia DR5, because RC magnitudes are brighter than the expected \textit{GaiaNIR} magnitude limit. We will therefore evaluate \textit{GaiaNIR} uncertainties in the mock catalogue using e.g. main-sequence turnoff stars as tracers in future work.


\section{Conclusions}
\label{sec:conclusions} 
In this work, we have presented a method to quantify the LMC kinematic asymmetries in the bar region and the outer disc previously noted by \citetalias{Jimenez-Arranz23a}. We analysed the KRATOS suite of $N$-body simulations \citepalias{Jimenez-Arranz24a} of LMC-like galaxies in isolation or in interaction with SMC- and MW-mass systems and found that the dynamical effect of the SMC can naturally cause asymmetries in the LMC kinematics. We also connected our analysis of the KRATOS simulations to \gaia observations of the LMC using the LMC optimal sample \citepalias{Jimenez-Arranz23a}. We explored possible causes of the asymmetries, including a bias in the LMC optimal sample classifier, and used our mock catalogue tool to evaluate the effects of the uncertainties from \gaia observational errors and the selection function. Lastly, we also considered the improvements of LMC kinematics studies with future releases of \gaia and \textit{GaiaNIR} data. The main findings of this work are the following: 
\begin{itemize}
    \item We have identified a velocity map of the K6 simulation at $t = 0.105$ Ga that better matches the features of the LMC optimal sample velocity maps (bottom row of Fig.~\ref{fig:velocity}). We recommend its use as the ``present time'' in future work with KRATOS.
    \item As expected, an isolated \lmc does not present the same asymmetries found in cases where the \smc is included (Figs.~\ref{fig:barevol} and~\ref{fig:discevol}).
    \item The velocity maps do not change when \smc stars are incorporated, suggesting that the effect of the \smc on the \lmc is purely dynamical (Fig.~\ref{fig:smc}).
    \item We investigated the observational uncertainties in \gaia data as a potential cause of kinematic asymmetries and found a bias in the LMC optimal sample classifier (Fig.~\ref{fig:NNbias}). We therefore conclude that the asymmetry in the outskirts of the disc observed in this sample is not intrinsic to the LMC but stems from this bias. 
    \item In KRATOS, asymmetries in the velocity maps of the bar and outer disc are generally correlated with pericentric passages, and closer pericentric passages cause larger variations in the asymmetry (Fig.~\ref{fig:allKRATOS}).
    \item Although we lack time evolution in the \gaia data, we can use our method to quantify the current asymmetry in the LMC optimal sample (Figs.~\ref{fig:barevol},~\ref{fig:discevol} and~\ref{fig:allKRATOS}). 
    \item From our study of mock catalogues we found that applying \gaia observational errors to the simulations does not significantly change the asymmetric signal in the bar quadrupole of the velocity maps (Fig.~\ref{fig:residuals_sf}). We conclude that this asymmetry in the LMC is due to its interaction with the SMC.
    \item The quadrupole signature is preserved independently of observational uncertainties, including the mock catalogues of future data releases \gaia DR4, \gaia DR5, and \textit{GaiaNIR} (Fig.~\ref{fig:residuals}). 
    
\end{itemize}

We therefore conclude that the asymmetries in the LMC bar of the \gaia velocity maps are a result of the interaction history between the LMC and the SMC, while the outer asymmetry observed in the LMC optimal sample is an effect of the sample classifier.

\begin{acknowledgements}
We thank the anonymous referee for providing constructive comments that improved the quality of this paper. We thank A. Castro-Ginard for insightful discussions and help in creating the \gaia selection function using \textsc{GaiaUnlimited}.
This work has made use of data from the European Space Agency (ESA) mission {\it Gaia} (\url{https://www.cosmos.esa.int/gaia}), processed by the {\it Gaia} Data Processing and Analysis Consortium (DPAC,
\url{https://www.cosmos.esa.int/web/gaia/dpac/consortium}). Funding for the DPAC has been provided by national institutions, in particular the institutions participating in the {\it Gaia} Multilateral Agreement. MS acknowledges funding by the European Union under the Horizon Europe Marie Skłodowska-Curie Actions Doctoral Network grant agreement no. 101072454 @HorizonEU research and innovation programme. OJA acknowledges funding from ``Swedish National Space Agency 2023-00154 David Hobbs The GaiaNIR Mission'' and ``Swedish National Space Agency 2023-00137 David Hobbs The Extended Gaia Mission''. MRG and XL acknowledge that this work was (partially) supported by the Spanish MICIN/AEI/10.13039/501100011033 and by "ERDF A way of making Europe" by the European Union through grant PID2021-122842OB-C21, and the Institute of Cosmos Sciences University of Barcelona (ICCUB, Unidad de Excelencia Mar\'{\i}a de Maeztu) through grant CEX2019-000918-M.
\end{acknowledgements}

\bibliographystyle{aa} 
\bibliography{bibliography}

\clearpage
\appendix

\end{document}